\documentclass[12pt]{article}

\usepackage{amsmath}
\usepackage{amssymb,amsthm,amsxtra}

\usepackage{epsfig}   

\usepackage{tikz-cd}
\newcommand{\ms}{\medskip}
\newcommand{\ve}{\varepsilon}

\newcommand{\ba}{\begin{array}}
\newcommand{\ea}{\end{array}}

\newcommand{\ds}{\displaystyle}

\newcounter{bean}
    {
      \begin{list}{\bf #1(\arabic{bean})}
         {\usecounter{bean}
              \setcounter{bean}{-1}
          \labelsep=1em
          \settowidth{\labelwidth}{#1\thebean:}
          \addtolength{\labelwidth}{1.1ex} 
          \leftmargin=\labelwidth 
          \addtolength{\leftmargin}{\labelsep} }

    }    {\end{list}}

\makeatletter
\@addtoreset{equation}{section}
\makeatother

\numberwithin{equation}{section}

\makeatletter
\def\iint{\DOTSI\protect\ints@\tw@}
\def\iiint{\DOTSI\protect\ints@\thr@@}
\def\iiiint{\DOTSI\protect\ints@{4}}
\def\idotsint{\DOTSI\protect\ints@\z@}

\def\intkern@{\mkern-6mu\mathchoice{\mkern-3mu}{}{}{}}
\let\DOTSI\relax
\let\ilimits@\displaylimits

\def\ints@#1{%
  \mkern-7mu\mathchoice{\mkern-2mu}{}{}{}%
  \mathop{\mkern7mu\mathchoice{\mkern2mu}{}{}{}%
    \intop\ifnum#1=\z@\intdots@
    \else\intkern@\fi
    \ifnum#1>\tw@\intop\intkern@\fi
    \ifnum#1>\thr@@\intop\intkern@\fi
    \intop
  }\ilimits@
}
\makeatother

\makeatletter
\def\@maketitle{\newpage
 \null
 \vskip 2em
 \begin{center}%
  {\normalsize\bf \@title \par}%
  \vskip 1.5em
  {\normalsize
   \lineskip .5em
   \begin{tabular}[t]{c}\@author
   \end{tabular}\par}%
  \vskip 2em
  {\@date}%
 \end{center}%
 \par
 \vskip 2.5em}
\makeatother

\makeatletter
\renewcommand\section{\@startsection {section}{1}{\z@}%
                                   {-3.5ex \@plus -1ex \@minus -.2ex}%
                                   {2.3ex \@plus.2ex}%
                                   {\normalfont\normalsize\bfseries}}
\renewcommand\subsection{\@startsection{subsection}{2}{\z@}%
                                     {-3.25ex\@plus -1ex \@minus -.2ex}%
                                     {1.5ex \@plus .2ex}%
                                     {\normalfont\normalsize\bfseries}}

\@addtoreset{equation}{section}
\makeatother

\begin{document}
\thispagestyle{empty}

\vspace*{3cm}

\begin{center}
\large{\bf Self--Organization,  Evolutionary Entropy and  Directionality Theory }

\vspace{1cm}
 \textbf{ Lloyd A. Demetrius}\\ 
  Dept. of Organismic and Evolutionary Biology,
  Harvard University\\ 
  Cambridge, Massachusetts 02138, U.S.A.\\
 April 10, 2023

\end{center}

\newpage
\begin{abstract}
Self--organization is the autonomous  assembly of a network  of interacting  components into a stable,  organized pattern. This article shows that the process of      self--assembly   can be  encoded  in terms of  evolutionary entropy, a statistical measure of the  cooperativity of the  interacting components. Evolutionary entropy describes the rate at which a  network of  metabolic components transduce an  external energy source into mechanical energy and work.

\ms
 We invoke \textit{Directionality  Theory,} an analytic model  of the collective behavior of a network of  interacting components, to show that the spontaneous emergence of organization can be depicted as the outcome of a  fluctuation--selection process, and articulated in terms of the following tenet. 
 
 \ms
\textit{The Entropic Principle of Self--Organization:} The equilibrium states of a self--assembly process are states which maximize evolutionary entropy, contingent on the production rate of the external energy source.

\ms
The Entropic Principle of Self--Organization is a universal rule, which pertains to the self--assembly of processes in various disciplines: Physics --- phase transitions;  Chemistry --- molecular assembly; Biology --- protein folding and morphogenesis; Sociology ---  the emergence of institutions.

\ms
The principle also elucidates the origin of cellular life; ---- the transition from inorganic matter to the emergence of protocells, capable of replication and metabolism.
 
 D
\end{abstract}
\newpage
\centerline{\large\bf Contents}
\begin{enumerate}
	\item[(1)] \textbf{Introduction}
	\begin{enumerate}
		\item[(1.1)]  \textit{Static Self--Assembly}
	\item[(1.2)] \textit{ Dynamic Self--Assembly}
	\item[(1.3)] \textit{ Self--Organization and Directionality Theory}
\item[(1.4)] \textit{ Organization of Article}
\end{enumerate}	
\item[(2)]\textbf{Thermodynamic Entropy and Evolutionary Entropy}
\begin{enumerate}
	\item[(2.1)] \textit{Thermodynamic Entropy}
	\item[(2.2)] \textit{ Evolutionary Entropy}
\begin{enumerate}
	\item[(2.2.1)] 	Age--structured populations	
	\item[(2.2.2)] Stage Structured populations
	\end{enumerate}
\end{enumerate}			
\item[(3)] \textbf{Evolutionary Entropy and the Thermodynamic Formalism}	
\begin{enumerate}
	\item[(3.1)]  \textit{ Thermodynamic Entropy ---  Physical Systems}
	\item[(3.2)] \textit{  Evolutionary Entropy --- Biological Systems}	
	\item[(3.3)]  \textit{Evolutionary Entropy and Modes of Interaction}  
 	\item[(3.4)] \textit{ Evolutionary entropy and Cooperativity}	
 \item[(3.5)]  \textit{Evolutionary entropy and Stability	}
 \item[(3.6)] \textit{Evolutionary Entropy and Darwinian Fitness}
\end{enumerate}	
\item[(4)] \textbf{Collective Behavior: Directionality Theory and Self--Organization }
\begin{enumerate}
	\item[(4.1)]  \textit{Directionality Theory and Collective Behavior}
	\item[(4.2)] \textit{ Self--Organization and Collective Behavior}	
	 \end{enumerate}	
\item[(5)] \textbf{Self--Organization: Static and Dynamic Self--Assembly}	
\begin{enumerate}
	\item[(5.1)]  \textit{Static Self--Assembly ---  Protein Folding Problem}
	\item[(5.2)] \textit{ Dynamic Self--Assembly ---  Benard Convection}	
\end{enumerate}		
\item[(6)] \textbf{Conclusion}
\begin{enumerate}
	\item[(6.1)]  \textit{Variation by mutation: The Evolutionary Process}
	\item[(6.2)]  \textit{Variation by fluctuation: The Self--Organization Process}
	\item[(6.3)] \textit{The steady State: Self--Organization and Evolutionary Processes} 	
\end{enumerate}
\end{enumerate}		

\newpage
\centerline{\large\bf 1. Introduction}

\medskip
Self--organization is a process of self--assembly in which stable, macroscopic structures spontaneously emerge as the outcome of the collective behavior of  components whose interaction is contingent on an external energy source.

\ms
The autonomous assembly of the interacting components can be explained in terms of a dynamical model which assumes that the macroscopic aggregates are composed of structurally unique components. The emergence of a stable, macroscopic pattern from a random assembly of interacting units can be articulated in terms of three principles
\begin{enumerate}
	\item [A(i)]\textit{ Variation: } The components  vary in terms of their cooperativity, that is their capacity to generate coherent, functional structures from local interactions. 
	\item [A(ii)] \textit{ Selection: } Local structures with a high degree of cooperativity have an enhanced capacity to appropriate and transduce an  external energy source into mechanical energy and work.
	\item [A(iii)] \textit{ Stability: } There is a positive correlation between the cooperativity of a local structure and its stability, that is the rate at which the structure returns to its steady state  after a  perturbation of its  components	
	\end{enumerate}
These principles imply that self--assembly is a hierarchical process, initiated by  structures  which are random in organization, and marginal in stability. The assembly is driven by a variation--selection process. Variation, the induction of diverse local structures,  is generated by fluctuations, an effect caused by the lability of the interactions between the components.Variation induces  \textit{reversible} changes in local structure.  Selection, the competition for energy, between local structures, is engendered  by differences in organization and   stability of the  variant structures.

\ms
The outcome of this fluctuation--selection process is the emergence of intermediates defined by increased cooperativity, and consequently, enhanced stability. The steady state  which this dynamic  process establishes  will be an organized pattern with maximal stability.

\ms
The  conservative  or dissipative nature of the energy source  which drives this process of organization can be invoked to distinguish between two mechanisms of assembly --- static and dynamic, Whitesides and Grzbowski (2002), Nicolis and Prigogine (1977), Lehn (2012), Haken (1977).

\bigskip
\textbf{(1.1) Static Self--Assembly}

\smallskip
Static self-assembly describes a process where there is no dissipation of energy. The organized pattern generated under this  energy constraint  will manifest a local or global equilibrium.

\ms
Cooperativity in systems generated by static self--assembly is generally based on physical interactions which implicate  parameters such as shape, charge, and polarizability. The components in these structures are held together by non--covalent intermolecular forces.

\ms
This process of self--assembly is manifest primarily at molecular and supramolecular scales. Static self--assembly can be illustrated by the following examples:

\begin{enumerate}
	\item[(a)] \textit{Molecular crystals: } Molecular crystals, for example, solid forms of noble gases and crystals of organic compounds are organic molecules whose structure is the outcome of static self--assembly. The organization of these molecules  is regulated by a balance between inter-- and intramolecular interactions, Lehn (2012).
	\item[(b)] 	\textit{The Folding of small  Proteins:} Protein folding is defined  by the transfer of a  one--dimensional sequence data into a three dimensional structure. The stabilization of the native state results from the balance of large internal forces that favor folding and disaggregation. These forces are due to non--covalent interaction forming oligomers of varying sizes and structure, Shakhnovich (1984).
\end{enumerate}	
		Static self--assembly at the organismic level is rare, Whitesides and Grzbowski (2002). An example of this mode of self assembly at  the organismic scale is aggregation in whirligig  beetles,  small immobile  aquatic organisms. The forces which ensure the emergence of stable patterns are due to capillary interactions, and the effect of surface tension, Voise et al. (2011).

\ms
\textbf{(1.2)  Dynamic Self--Assembly}

\smallskip
Dynamic self--assembly describes a process in which the aggregates change and evolve only if energy is continuously delivered to the components. The equilibrium configuration which this process induces is a steady state. The pattern of organization, its architecture and stability, depends on the amount of energy delivered to the system.

\ms
Dynamic self--assembly is manifest at all scales: molecular, cellular, social and economic, Fialkowski et al. (2005), Camazine (2003), Saha and Galic (2018). We will illustrate the phenomena  by examples drawn from two levels of  organization: cellular (Benard convection), molecular (microtubular morphogenesis).

\bigskip
\textbf{(a)} \textit{Benard Convection:}

\smallskip
Benard cells form when a viscous fluid is heated between two  planes in a gravitational field. The formation of the geometric pattern  --- the Benard cells --- depends on the type of fluid, its depth and the temperature gradient, Nicolis and Prigogine (1977).

\ms
The induction of the  geometric pattern is contingent on the relation between  fluctuations in the density of the fluid, and its viscosity. There is a critical parameter  at which fluctuations in fluid density  overcome the viscosity faster than they are dissipated. At this juncture, the  patterns induced by the fluctuations  are amplified, and generate a macroscopic geometric current --- the Benard cells. The critical parameter at which the transition to macroscopic order occurs, depends on the relation between the  gravitational force, and the viscous damping force in the fluid.

\bigskip
\textbf{(b)}\textit{ Microtubules and Morphogenesis}

\smallskip
A standard example of self-organization at the molecular level  is the phenomenon of microtubular  morphogenesis --- the spontaneous assembly of long uniform polymers from actin and tubulin subunits, Kirschner and Mitchison (1961), Tabony et al. (2022).

\ms
The elements actin  and tubulin polymerize to form filaments and cytoskeletal microtubules that define the shape of a cell --- its migration and polarity. Although microtubule assembly bears a resemblance  to  well studied systems such as virus assembly, it possesses  several important differences. First, the assembly of microtubules requires an external energy source.  Second, polymer assembly is regulated.  In freely growing microtubules, the stability of the polymer is controlled by the presence or absence of subunits containing unhydrolyzed GTP.

\ms
Empirical studies  of the morphogenetic process show that  the transition from a disordered population to an organized state occurs  at a critical concentration of tubulin.

\bigskip
\textbf{1.3 Self--Organization and Directionality Theory}

\small
Static and dynamic assembly are processes which manifest at all scales of organization: molecular, cellular and multicellular, Whitesides and Grzlowski (2002), Camazine (2003), Karsento (2008), Saha and Galic (2018). The phenomenon of self--assembly is also observed in social and economic systems, Krugman (1996).

\ms
The problem  of self--organization can be formalized in terms of the following query.  Can the emergence of organization, a derivative of the dynamic interaction between external forces and local constraints, be encoded in terms of general principle of self--assembly?

\ms
The problem has a long history, and partial answers have been proposed based on studies at particular scales of organization. Important contributions towards the resolution of the problem include the reaction--diffusion models of morphogenesis, Turing (1952); the studies based on non--equilibrium thermodynamics, Prigogine and Nicolis (1977); and the analysis based on adaptive chemistry, Lehn (2012).

\ms
The analysis  of Self--organization invoked in this article is based  on\textit{ Directionality Theory,} the study of the collective behavior, due to variation and selection, of interacting components, Demetrius (1997), (2013).  The central concept of Directionality Theory is evolutionary entropy,  a generalization of the thermodynamic entropy of Gibbs and Boltzmann. Evolutionary entropy This concept is a statistical measure of cooperativity, which we define as the extent to which the individual components of the network are held together by the interacting forces. These forces may be intermolecular, modulated by molecular recognition;    intercellular,  encoded in terms of  physiology; or organismic, coordinated by individual behavior.

\ms
Evolutionary entropy describes the rate at which the network of interacting components transduce  resources into mechanical energy and work. Accordingly, evolutionary entropy determines the outcome of selection, that is competition between variant networks for the  external energy source, Demetrius (2013), Demetrius, Gundlach and Ochs (2009).

\ms
We will consider  a Self--Organizing process as an instance of collective behavior. Self--assembly is assumed to evolve, by fluctuation and selection, through a succession of hierarchical transitions,   whereby local structures cohere, and thereby generate intermediates of increasing complexity and stability.

\ms
We will show that the steady state induced by this fluctuation--selection  process can be articulated in terms of  the following rule.

\ms
\textit{The Entropic Principle of Self--Organization:} The equilibrium states of a self--assembly process is contingent on the external resource constraints, and characterized by the maximization of evolutionary entropy.

\ms
The analytical expression of the rule is the relation
$$
-\Phi\Delta H \ge 0\leqno(1)
$$
The expression $\Phi$, called the organizing potential,  encodes the production rate of the energy source. The function $\Delta H = H^*-H$, where $H$ denote the evolutionary entropy of an ancestral structure, and $H^*$ the evolutionary entropy of a variant structure.

\ms
The relation between the organizing potential $\Phi$ and the change $\Delta H$ in evolutionary entropy is depicted in Table (1)

\ms
\begin{center}
Table (1): Relation between the organizing potential and  change in evolutionary entropy

\ms
\begin{tabular}{|l|l|}
	\hline
	\textbf{Organizing Potential}&\textbf{Changes in Evolutionary Entropy}\\
	$\Phi < 0$&$\Delta H > 0$\\
	$\Phi > 0$&$\Delta H < 0$\\
	\hline
\end{tabular}
\end{center}

\ms
We will show that the expression, given by (1), is a limiting case of a more general tenet, the \textit{Entropic Principle of Evolution.}

\ms
The Entropic Principle of Evolution describes the steady state of \textit{Directionality Theory.}

\ms
The changes in cooperativity of the interacting components which characterizes the evolutionary process in Directionality Theory, are also induced by a variation--selection process. Variation, in this context, pertains to \textit{irreversible} changes in local structure. 

\ms
The fundamental unit of Directionality Theory is a population, which consists of individuals with metabolic and replicative capacities. The model assumes that the individuals in the population interact by appropriating energy from the external environment.  Structural and cooperative units are formed through the sharing and distribution of energy among the individuals.

\ms
The adaptation of the population to the  environmental constraints can be expressed in terms of the following  principles.
\begin{enumerate}
	\item [(1)]\textit{ Variation: } The components that comprise the population vary in terms of their physical and biological properties.
	\item [(2)] \textit{ Succession: } There exists a positive correlation between the properties of a component and properties of its replicas.
	\item [(3)] \textit{ Selection: } The external energy source vary in terms of its amplitude and constancy. Components with different physical and biological properties vary in terms of their capacity to transduce  the external energy into other forms of energy and work.
\end{enumerate}

These principles entail that changes in the composition of the population  will occur as one variant population  replaces another due to competition for the external energy source.

\ms
These changes in composition are parametrized in terms of evolutionary entropy, and are  expressed in terms of the following rule 

\ms
\textit{The Entropic Principle of Evolution:  } The equilibrium states of the evolutionary process of variation and selection are contingent on the external energy source  and population size, and characterized by extremal states of evolutionary entropy.

\ms
The analytical encoding of the Entropic Principle of Evolution is the relation,
$$
(-\Phi+\gamma/M)\Delta H \ge 0\leqno(2)
$$
The function $\Phi$ represents the resource production rate;  $\gamma$ denote the temporal correlation in resource production, $M$ the total population size. The function $\Delta H = H^*-H$.

\ms
The relation between the changes $\Delta H$ in evolutionary entropy, and the parameters $\Phi$, $\gamma$ and the population size $M$ are depicted in Table (2) and Table (3).

\ms
Table (2) describes the relation between the resource constraints and changes in evolutionary entropy, when $\Phi$ and $\gamma$ are negatively correlated, ($\Phi\gamma < 0$)

\ms
Table (3) describes the relation between the resource constraints and changes in evolutionary entropy, when $\Phi$ and $\gamma$ are positively correlated, ($\Phi\gamma > 0$)

\ms
\begin{center}
	Table (2): Directional Changes in Evolutionary Entropy: ($\Phi\gamma < 0$)
	
	\ms
	\begin{tabular}{|l|l|}
		\hline
		\textbf{Constraints on the parameters} &\textbf{Directional Changes in Evolutionary}\\
		\textbf{$\Phi$ and  $\gamma$}&\textbf{ Entropy}\\
	\hline
		$\Phi < 0$, $\gamma  > 0$& $\Delta H > 0$\\
		$\Phi > 0$, $\gamma  < 0$&   $\Delta H < 0$\\
		\hline
	\end{tabular}

\pagebreak
Table (3): Directional Changes in Evolutionary Entropy: ($\Phi\gamma > 0$)

\ms
\begin{tabular}{|l|l|}
	\hline
	\textbf{Constraints on $\Phi, \gamma$}&\textbf{Directional changes: $\Delta H$} \\
	\hline
$\Phi < 0, \, \, \gamma < 0$& \\
	\quad $\gamma > M\Phi$&$\Delta H > 0$\\
	\quad $\gamma < M\Phi$&$\Delta H < 0$\\
	$\Phi > 0,\, \, \gamma > 0$& \\
	\quad $\gamma > M\Phi$&$\Delta H > 0$\\
	\quad $\gamma < M\Phi$&$\Delta H < 0$\\       
	\hline
\end{tabular}
\end{center}

\ms
Directionality Theory is an abstract model of the evolution and adaptive dynamics of the collective behavior of matter, inorganic and organic, due to variation and natural selection; Demetrius (2013).  The theory has been applied in various contexts, namely: the evolution of life history, Demetrius (2013), the evolution of social behavior, Demetrius and Gundlach (2015), the evolution of economic inequality, Germano (2022).

\ms
The analytic formalism invoked in the different applications is dictated by the mechanism which drives the process of variation. This process, in studies of the evolution of life--history, is determined by \textit{mutations,} changes in the genes that encode individual survivorship and reproduction. These changes are \textit{irreversible}. The process of mutation will induce absolute changes in demographic states due to the  uni--directional flow, DNA $\to$ RNA $\to$ Protein, that defines  cellular life.

\ms
In this article, we will apply the formalism of Directionality theory to analyze the adaptive dynamics of Self--organization. The variation process in Self Organization,  the autonomous assembly of interacting components, is \textit{fluctuations, } a reorganization of the individual components. These changes are \textit{reversible.}  The reversibility is due to the lability of the interacting components. We will show that the changes in evolutionary entropy induced by the fluctuation--selection process that defines Self--Organization are a derivative of  Eq. (2), and is given by Eq. (1).

\ms
The relation between Directionality theory, a general model of the evolution of collective behavior under variation and selection, and Self--Organization, the evolution of collective behavior due to fluctuation and selection, is described in Fig. (1)

\ms
\begin{center}
	\includegraphics[width=5cm]{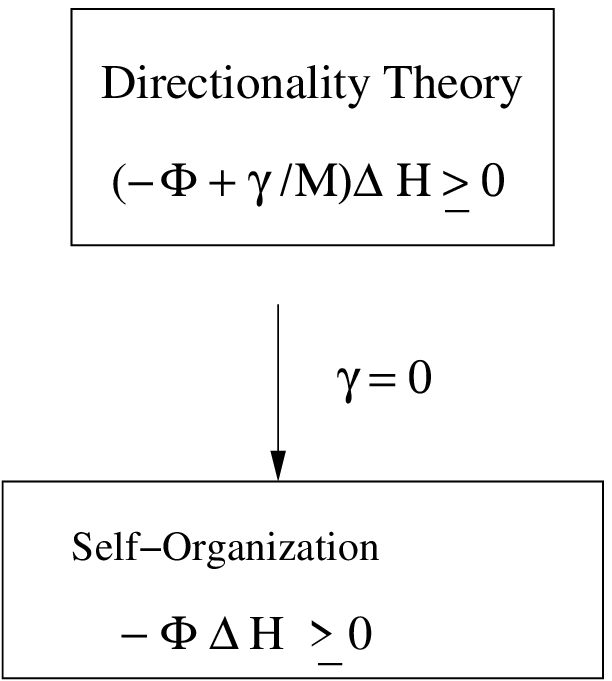}
	
	\ms
	\textbf{Fig. (1)} Relation between Directionality Theory and Self--Organization.
\end{center}

\bigskip
\textbf{1.4 Organization of article}

\smallskip
This article is organized as follows: The measure of cooperativity invoked in Directionality theory is evolutionary entropy, a generalization of the thermodynamic entropy of Gibbs and Boltzmann.

\ms
Section 2. reviews the mathematical basis of  evolutionary entropy,  and elucidates its relation with thermodynamic entropy.   
The properties of evolutionary entropy, as a measure of cooperativity, stability and Darwinian fitness, are reviewed in Section (3).\\
Section 4  gives a succinct account  of the mathematical basis of the \textit{Entropic Principle of Evolution,} the cornerstone of Directionality Theory. This section invokes Directionality Theory to derive the \textit{Entropic Principle of Self--Organization.}\\
Section 5   illustrates the Entropic Principle of Self--Organization by an  analysis of  Protein Folding, an example of Static Self--Assembly, and the Benard Convection cell, an example of Dynamic Self--Assembly. The Conclusion,  Section 6  explicates  the relation between the two theories, Self--Organization and Directionality Theory.

\newpage

\newpage
\centerline{\large\bf 2. Cooperativity: Thermodynamic Entropy and Evolutionary Entropy}

\ms
The fundamental unit in the study of the dynamics of collective behavior is a population. This object is defined in terms of a set of interacting components: molecules, macromolecules, cells, higher organisms. The components  are parametrized in terms of their physical or biological properties  --- the  microstates. We consider the population as a strongly connected, weighted, directed graph as in Fig. (2).  The nodes of the graph correspond to the set of microstates. The links between the nodes represent the transfer or flow of energy between the microstates.

\ms
\begin{center}
	\includegraphics[width=8cm]{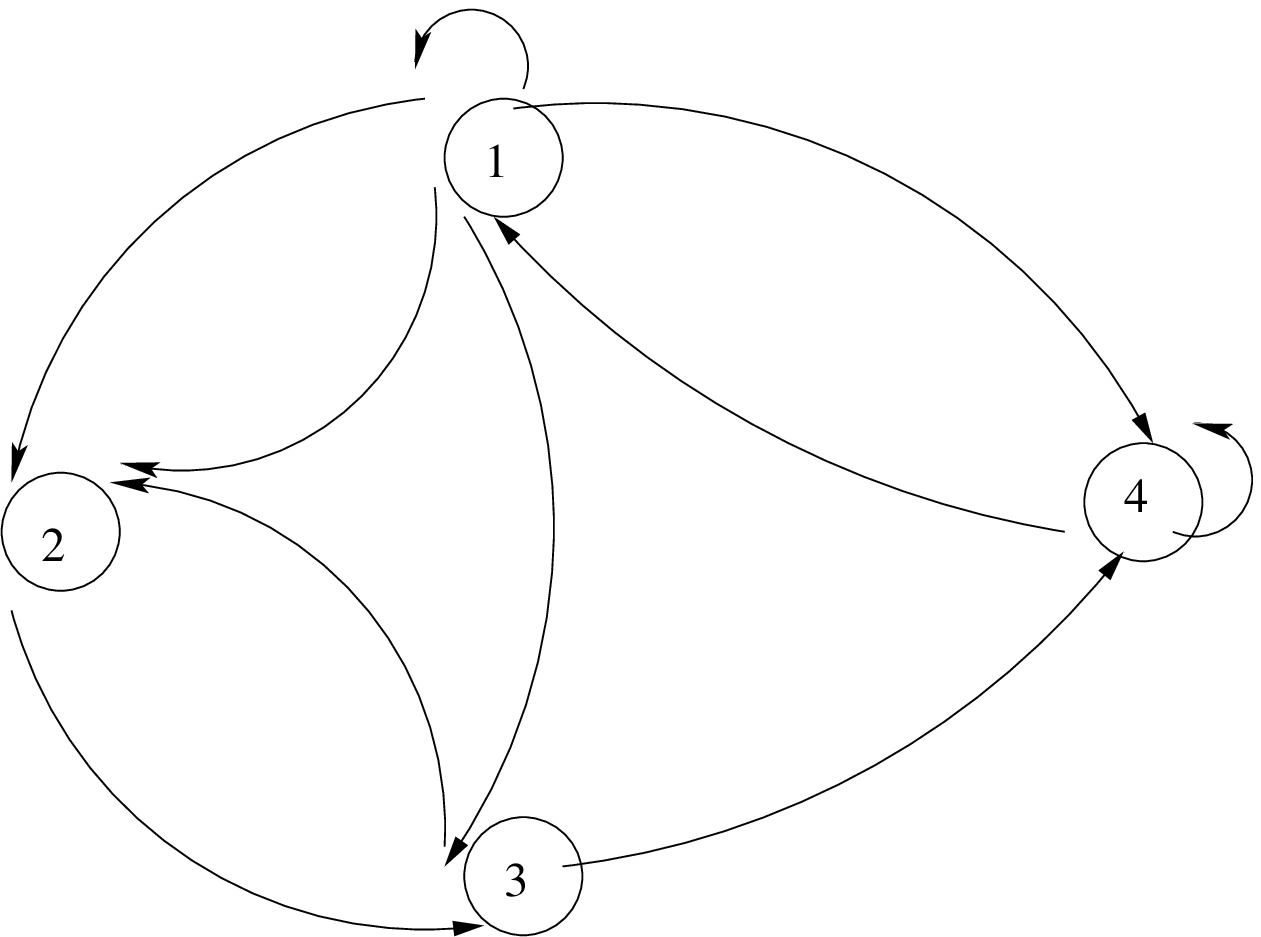}
	
	\ms
	\textbf{Fig. (2)} A network consisting of interacting components
\end{center}

\bigskip
\textbf{2.1 Thermodynamic Entropy:}

\smallskip
The interacting components in Statistical Thermodynamics are the atoms and molecules which comprise inanimate matter --- solids, liquids and gases. Each molecule in a solid stores translational, rotational and vibrational energy. Energy is spread throughout the macroscopic body as a result of the interaction between the indivdual molecules.

\ms
The analysis of the spreading and sharing of energy in the macroscopic body is based on the following assumptions:
\begin{enumerate}
	\item[(i)]  The number of molecules that comprise the aggregate is large, effectively infinite
	\item[(ii)]  The forces which determine the interaction between the molecules are short--ranged.
\end{enumerate}
Collective behavior of the components that comprise the macroscopic aggregate is described by  thermodynamic entropy, expressed by 
$$
S_B=k\log W\leqno(3)
$$
The quantity $W$ denote the total number of microstates of a system consistent with a given macrostate. 

\ms
If we assume that the particles in the macroscopic aggregate occupy small volumetric cells, $i=1,2,\ldots s$ of phase space, with occupation numbers $n_i$, and that the total number $N$ of particles is conserved, then
$$
N=\sum^s_{i=1} n_i
$$
The quantity $W$,  the number of instantaneous microstates is given by
$$
W=\frac{N!}{n_1!n_2!\ldots n_s}
$$
Assuming $N$ is large, we obtain from (3), by applying Stirling's approximation, the relation
$$
S_B=NS_G
$$
where
$$
S_G=-\sum_{j=1}\tilde{p}_j\log\tilde{p}_j
$$
where $\tilde{p}_j=\frac{n_j}{N}$. The element $\tilde{p}_j$ is the probability that a particle is in cell $(j)$, provided $N$ is sufficiently large.

\ms
The quantity $S_G$ is called the Gibbs entropy. Fig. 3(a) and Fig. 3(b) represent the distribution of particles in a solid and a gas, respectively.

\ms
\begin{center}
	\begin{tikzpicture}
	\draw[thick](0,0) rectangle (4,3);
	\node at (1.8,-0.6) {Solid};
	\node at (8.6,-0.6) {Gas};
	\draw[thick](6.5,0) rectangle (10.5,3);
	\filldraw[blue] (0.5,0.5) circle (0.075); \filldraw[blue]  (1,0.5) circle (0.075); \filldraw[blue]  (1.5,0.5) circle (0.075); \filldraw[blue]  (2,0.5) circle (0.075);\filldraw[blue]  (2.5,0.5) circle (0.075);\filldraw[blue]  (3,0.5) circle (0.075);\filldraw[blue]  (3.5,0.5) circle (0.075);
	\filldraw[blue]  (0.5,1) circle (0.075); \filldraw[blue]  (1,1) circle (0.075); \filldraw[blue]  (1.5,1) circle (0.075); \filldraw[blue]  (2,1) circle (0.075);\filldraw[blue]  (2.5,1) circle (0.075);\filldraw[blue]  (3,1) circle (0.075);\filldraw[blue]  (3.5,1) circle (0.075);
	\filldraw[blue]  (0.5,1.5) circle (0.075); \filldraw[blue]  (1,1.5) circle (0.075); \filldraw[blue]  (1.5,1.5) circle (0.075); \filldraw[blue]  (2,1.5) circle (0.075);\filldraw[blue]  (2.5,1.5) circle (0.075);\filldraw[blue]  (3,1.5) circle (0.075);\filldraw[blue]  (3.5,1.5) circle (0.075);
	\filldraw[blue]  (0.5,2) circle (0.075); \filldraw[blue]  (1,2) circle (0.075); \filldraw[blue]  (1.5,2) circle (0.075); \filldraw[blue]  (2,2) circle (0.075);\filldraw[blue]  (2.5,2) circle (0.075);\filldraw[blue]  (3,2) circle (0.075);\filldraw[blue]  (3.5,2) circle (0.075);
	\filldraw[blue]  (0.5,2.5) circle (0.075); \filldraw[blue]  (1,2.5) circle (0.075); \filldraw[blue]  (1.5,2.5) circle (0.075); \filldraw[blue]  (2,2.5) circle (0.075);\filldraw[blue]  (2.5,2.5) circle (0.075);\filldraw[blue]  (3,2.5) circle (0.075);\filldraw[blue]  (3.5,2.5) circle (0.075);
	\draw [thick,->](6.9,1.2) -- (6.9,1.6);
	\draw [thick,->](7.9,0.4) -- (7.6,0.9);
	\draw [thick,->](7,0.5) -- (7.3,0.8);
	\draw [thick,->](8,1.5) -- (7.7,1.2);
	\draw [thick,->](9,2.3) -- (9.5,2);
	\draw [thick,->](8.8,2.6) -- (9.2,2.7);
	\draw [thick,->](8.5,0.5) -- (9,0.6);
	\draw [thick,->](10,0.5) -- (9.8,1.2);
	\draw [thick,->](9.4,1.1) -- (9.7,0.8);
	\draw [thick,->](9,1.5) -- (9,1.1);
	\draw [thick,->](10.2,1.5) -- (9.6,1.5);
	\draw [thick,->](7,0.5) -- (7.3,0.8);
	\draw [thick,->](8,1.9) -- (7.6,1.9);
	\draw [thick,->](7.2,2.7) -- (7.1,2.3);
	\draw [thick,->](8,2.5) -- (7.4,2.4);
	\draw [thick,->](8.3,1.5) -- (8.3,1.8);
	\draw [thick,->](8.3,0.9) -- (8.6,1.3);
	\draw [thick,->](8.7,2.2) -- (8.6,2.8); 
	\draw [thick,->](9,1.8) -- (8.8,1.6);
	\draw [thick,->](9.9,2.7) -- (9.6,2.6);
	\draw [thick,->](9.6,2.1) -- (9.9,2.3);
	\draw [thick,->](9.9,2.7) -- (9.6,2.6);
	\draw [thick,->](7.3,1.8) -- (7.6,1.6);
	\draw [thick,->](9.4,0.3) -- (9.3,.6);
	\draw [thick,->](8,2.3) -- (8.3,2.2);
	\end{tikzpicture}
\end{center}
\centerline{ Fig. 3 (a)\qquad\qquad\qquad\qquad\qquad\qquad  Fig. 3 (b) }

\ms
The thermodynamic entropy will be small in the case of a solid, and large in the case of a gas.

\bigskip
\textbf{2.2 Evolutionary Entropy}

\smallskip
The microscopic components in Directionality Theory are  macromolecules, cells and higher organisms. The analysis of the spreading and sharing of energy in this class of populations is based on the following set of assumptions.
\begin{enumerate}
	\item[(i)]  The number of microstates that define the population is finite.
	\item[(ii)] The forces which determine the interaction between the microstates are long--ranged.	
\end{enumerate}
Collective behavior of the components that comprise the population is now described by the statistical parameter, evolutionary entropy $H$, given by $H=S/T$.

\ms
The quantity $S$ denote the number and diversity of interaction cycles generated by the directed graph. The quantity $T$ denote the  cycle time.

\ms
We illustrate the notion of evolutionary entropy by considering two classes of populations:
\begin{enumerate}
	\item[(i)] A population structured in terms of age.
	\item[(ii)]  A population structured in terms of some physiological or morphological property. 
	\end{enumerate}
 A formal analysis of the mathematical basis of evolutionary entropy is described in Section 3.2.
\bigskip
\textbf{2.2.1} \textit{Age--structured populations.}

\smallskip
Individuals in the population are classified in terms of their age. The microstates are age--classes. 
Individuals appropriate energy from the external environment and convert this energy into metabolic energy and biomass. Energy is spread throughout the population as a result of the processes of aging and reproduction.

\ms
The aging process induces the transfer of energy from age class (i) to age--class (i+1). The reproductive process is characterized by the transfer of energy from age--class (i) to age class (1), the microstate of newborns.

\ms
Collective behavior is described by the statistical parameter, evolutionary entropy, expressed by $H=S/T$.

\ms
The functions $S$ and $T$ are given by
$$
S=-\sum^d_{j=1}p_j\log p_j\, \, ,\quad T=\sum^d_{j=1}j\,  p_j\leqno(4)
$$
The function $p_j$ denote the probability that the mother of randomly chosen newborn belongs to age-- class (j).

\ms
Fig. 4(a) and 4(b) represent the age--specific survivorship and fecundity of a semelparous and iteroparous population respectively. In Fig. 4(a), evolutionary entropy $H=0$. In Fig. 4(b), $H > 0$. In Fig. 4(a), the individuals in the population reproduce at a single instant in their life--cycle. In Fig. 4(b) reproduction occurs at several distinct stages in the life--cycle.

\ms
\begin{center}
	\includegraphics[width=10cm]{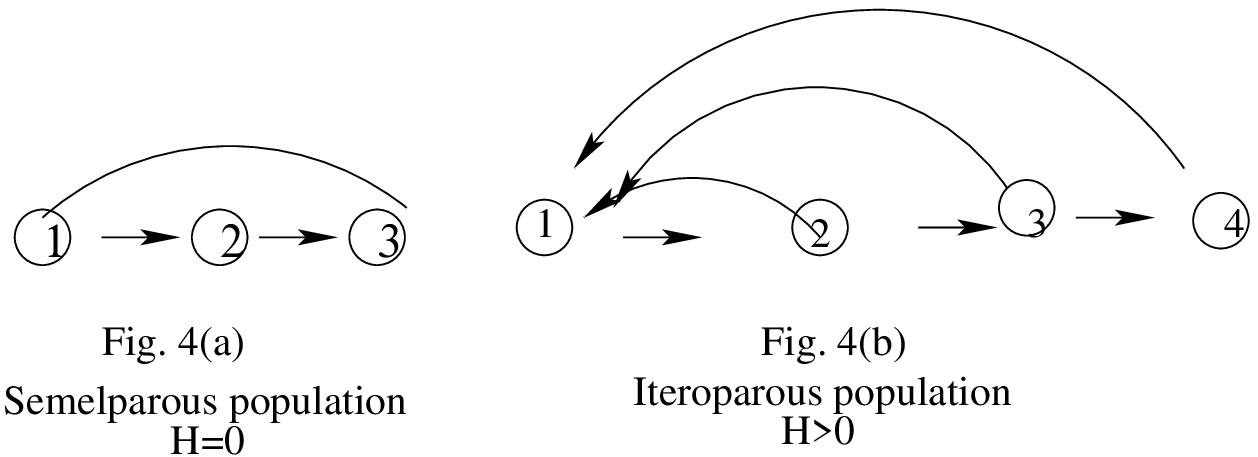}
\end{center}

\bigskip
\textbf{2.2.2} \textit{Stage--Structured population}

\smallskip
Individuals are classified in terms of a morphological or physiological trait. Energy is spread throughout the population by the transfer of resources between the different microstates --- the morphological or phyological states in which the individuals are classified.

\ms
Collective behavior of the stage--structured population  is described by evolutionary entropy. The quantity $S$ denote the number and diversity of interaction cycles generated by the directed graph. $T$ denotes the mean cycle time of the process.

\ms
The quantities $S$ and $T$ are described as follows.

\ms
We write
$$X=\{(1,2,\ldots,d)\}$$
This set $X$ describes the nodes of the graph. We now fix an arbitrary vertex $\alpha \in  X$ and we denote by $X^*$, the set of directed paths which starts at $\alpha$ and ends at $\alpha$, without traversing the node $\alpha$ in the middle.

\ms
An element $\tilde{\alpha} \in X^*$, is a sequence
$$
\alpha\to \beta_1 \to \beta_2\to \cdots \to \beta_{n-1}\to \alpha
$$ 
which we denote by
$$
\tilde{\alpha}=[\alpha\beta_1 \beta_2 \cdots  \beta_{n-1} \alpha].
$$
Consider the function 
$$
p_{\tilde{a}}=p_{\alpha\beta_1}p_{\beta_1\beta_2}\cdots p_{\beta_{n-1}\alpha}
$$ 
The quantity $T$, the cycle time, and $S$, the conformational entropy, are 
$$
T=\sum_{\tilde{\alpha}\in X^*}|\tilde{\alpha}|p_{\tilde{a}}\, ; \quad  S=-\sum_{\tilde{\alpha}\in X^*}p_{\tilde{a}}\log p_{\tilde{a}}
$$
Fig. 5(a), 5(b) and 5(c) represents systems with increasing values for the evolutionary entropy $H$. In Fig. 5(a), the evolutionary entropy assumes its minimum value. The maximum value for evolutionary entropy is achieved in Fig. 5(c)

\ms
\begin{center}
	\includegraphics[width=10cm]{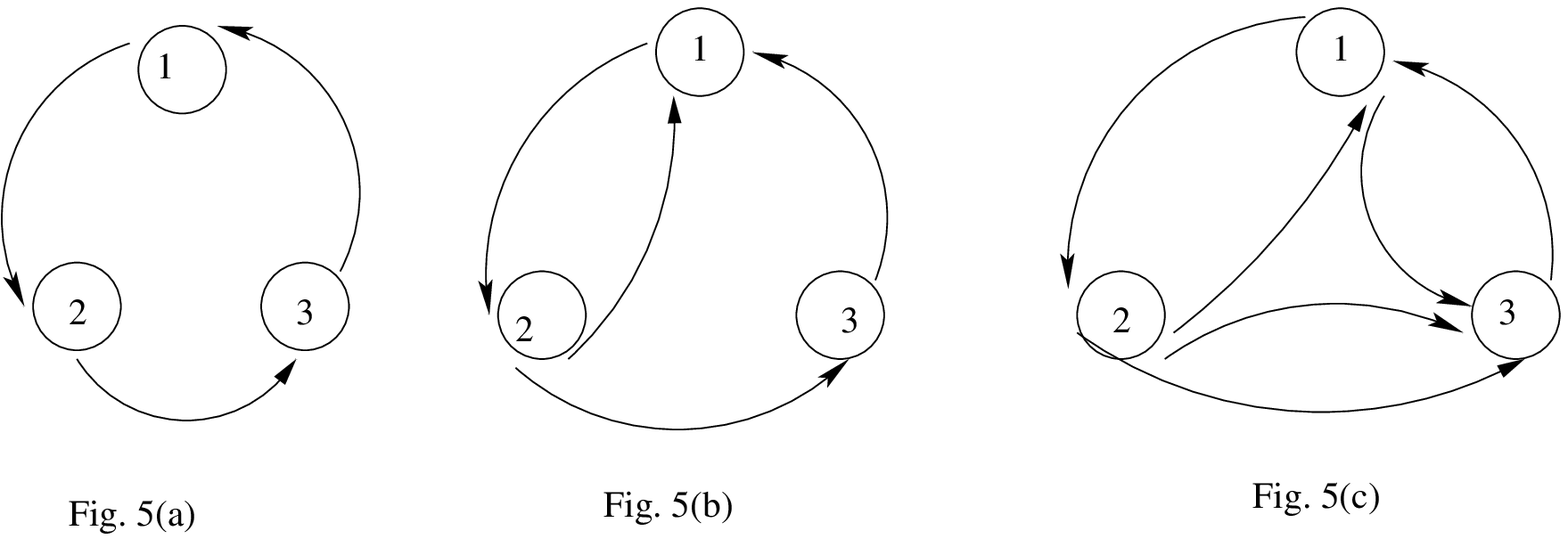}

\end{center}

\newpage

\centerline{\large\bf{3. Evolutionary Entropy and the Thermodynamic Formalism}}

\smallskip
Boltzmann and Gibbs developed a class of algorithms which furnished methods for deriving the macroscopic properties of an ideal gas from the variables which define the interaction between the individual atoms and molecules.

\ms
Collective behavior of the components that comprise the macroscopic aggregate is described in terms of the following  parameters: the free energy $F$; the mean energy  $E$, the thermodynamic entropy $S$, and the temperature $T$. These quantities satisfy the identity
$$
F=E-kST\leqno(5)
$$
We will exploit the methodology invoked in the derivation of (5) to generate a set of macroscopic variables to describe the collective behavior  of components which are defined in terms of long range interactions.

\ms
We first describe the derivation of the relation (5) for physical systems with finite phase space. Cooperativity is characterized by thermodynamic entropy.

\bigskip
\textbf{3.1.  Thermodynamic entropy --- Physical Systems}

\smallskip
We assume that the phase space of the system is finite, and  described by the mathematical object $(\Omega,\mu,\varphi)$, whose elements are characterized as follows
\begin{enumerate}
	\item[(i)] $\Omega$: The phase space.   The set $\Omega$ is the object $\Omega=(1,2,\ldots d)$
	\item[(ii)]	$\mu$:  A state, a  probability measure on the phase space  $\Omega$. 
	\item[(iii)]	$\varphi$: A potential function. This  function associates with each element in the phase space, a real number which represents the intensity and the nature of the interaction between the components.
\end{enumerate}
The model assumes that the forces which determine the interaction between the components of the system are short--ranged. The measure of cooperativity of the interacting components will be given by the Shannon entropy, denoted $S$, and given by
$$
S=-\sum^d_{i=1} \mu_i\log \mu_i\leqno(6) $$
The macroscopic parameters are
\begin{enumerate}
	\item[(i)] The mean energy $\Phi$ is given by
	$$
	\Phi=\sum^d_{\i=1}\mu_i\log \varphi_i\equiv \mu(\log \varphi)\leqno(7) 
	$$
\item[(ii)]		The free energy $F$ is given by
	$$
	F=\log Z
	$$
	where 
	$$
	Z=\sum^d_{i=1} \varphi(x_i)
	$$
	\end{enumerate}
The following statement is an immediate consequence of the definition of $S$, $\Phi$ and $\log Z$.

\ms
\textbf{Proposition (1): } $\log Z = \sup\limits_{\mu\in M}[S(\mu)+\mu(\log\varphi)]
$

\ms
The relation
$$
\log Z = S(\hat{\mu})+ \hat{\mu}(\log \varphi)\leqno(8)
$$
is attained by a unique $\mu=\hat{\mu}$, where
$$
\hat{\mu}=(\hat{\mu}_i) , \, \hat{\mu}_i=\frac{\varphi(x_i)}{Z}
$$
The variational principle which is formalized by Proposition (1) admits a physical interpretation. To observe this interpretation, we write $\varphi(x_i)= \exp(a_i)$, where $a_i=-\beta E_i$.

\ms
We consider the case, $\beta = 1/\kappa\, T$; where $T$ is temperature and $\kappa$ a physical constant.

\ms
Then the distribution $\hat{\mu} = (\hat{\mu}_j)$ is given by
$$
\hat{\mu}_j=\frac{\exp(-\beta E_j)}{\sum \exp(-\beta E_j)}$$
This is called the Gibbs distribution. This distribution maximizes
$$
S-\beta E=S-(\frac 1{\kappa T})E
$$
Equivalently, it minimizes 
$$
E-\kappa ST\leqno(9)
$$
We can describe the analogue of (9) for systems whose phase space is infinite. Cooperativity is now characterized by evolutionary entropy.

\bigskip
\textbf{3.2  Evolutionary Entropy --- Biological Systems}

\smallskip
We consider a population of interacting organic components as a dynamical system. We assume that the population is at steady state and that its collective behavior is described by the dynamical system $(\Omega,\mu,\varphi)$, whose elements are described as follows:
\begin{enumerate}
	\item[(i)] $\Omega$: The phase space of the dynamical system. $\Omega$ is the set of paths generated by the interaction between the components of the network.
	\item[(ii)]	$\mu$:  A state. A  probability measure on the phase space  $\Omega$. 
	\item[(iii)]	$\varphi$: A potential function. The function associates with each element in the phase space, a real number which represents the intensity and the nature of the interaction between the components.
\end{enumerate}
Consider the set $X=(1,2,\ldots d)$,  as representing the set of nodes of the graph.

\ms
The elements of the mathematical object $(\Omega,\mu,\varphi)$ can be characterized as follows. Write
$$
Y=\prod\limits^\infty_{i=1} X_n
$$
where
$$
X_n=X
$$
Let $A=(a_{ij})$ be a $d\times d$ matrix  with entries in $(0,1)$.

\ms
The phase space $\Omega$ can be described by the object 
$$
\Omega=\{x\in Y : a_{x_kx_{k+1}}=1\}
$$
An element $x\in \Omega$ is described by
$$
x=(\ldots x_{-1}x_0,x_1\ldots )
$$
where $x_i\in X$.

\ms
The element $x\in \Omega$ is called a genealogy. It specifies a path of the graph, which is a sequence generated by the interaction between the components that define the network.

\ms
The potential $\varphi$ is a rule which associates with each genealogy, a real number which encodes the interaction between the components that comprise the genealogy.

\ms
The forces which determine the interaction between the components of the network are long--range. The steady state will be represented by the dynamical system $(\Omega,\mu,\tau)$.

\ms
The parameter $\tau$ is the shift operator on $\Omega$ defined by 
$$
\tau:(x_\kappa)\mapsto (x'_\kappa)
$$
where $x'_\kappa=x_{\kappa+1}$.

\ms
Cooperative behavior in the dynamical system will be described by the dynamical entropy $H_\mu(\tau)$.

\ms
The dynamical entropy, a generalization of the Shannon entropy, $S(\mu)$, is an entropy rate, a quantity which describes the transmission of information in probabilistic dynamical processes. Dynamical entropy $H_\mu(\tau)$ has the physical connotation of mean information transmission per unit time, Bowen (1975). It is an isomorphism invariant of the dynamical system. Two dynamical systems are said to be isomorphic if and only if one can find a one--to--one correspondence between all (but a set of measure zero) of the  points in each configuration space so that the corresponding points are transformed the same way.

\ms
The dynamical entropy $H_\mu(\tau)$ is a fundamental property of the dynamical system $(\Omega,\mu,\tau)$.

\ms
The macroscopic parameters that describe the collective behavior of the population are analogues of the mean energy, the free energy, and the temperature.

\ms
The reproductive potential, denoted $\Phi$, the analogue of the mean energy, is given by 
$$
\Phi=\int \varphi\, d\mu$$
The population growth rate, the analogue of the free energy, is described as follows.

\ms
Write
$$
S_m(\varphi) = \sum^{m-1}_{\kappa=0} \varphi(\tau^\kappa x)
$$

\ms
Define 
$$
Z_m(\varphi)=\sum\limits_{x_0x_1\ldots x_n}[\exp(S_m\varphi(x^*)]
$$
where, for given $(x_0,x_1\ldots x_m)$, we denote $x^*$ any points in $\Omega$, with $x^* _i=x_i$ for $i=0,1,2,\ldots m$. 

\ms
The growth rate $r(\varphi)$ defined by 
$$
r(\varphi)= \lim\limits_{m\to\infty}\left[\frac 1m\right] \log  Z_m(\varphi)
$$
The quantity $r(\varphi)$ satisfies a variational principle, Bowen (1975), Ruelle (1974)
$$
r(\varphi)= \sup\limits_{\mu\in M}\left[H_\mu(\tau)+\int \varphi dm\right]
$$
The expression $M$ denotes the set of probability measures on $\Omega$, which are invariant under the shift $\tau$.

\ms
We define an equilibrium state of the system $(\Omega,\mu,\varphi)$ as the probability measure $\mu$, which satisfies the condition
$$
r(\varphi)=H_{\hat{\mu}}(\tau) + \int \varphi d\hat{\mu}\leqno(10)
$$
Let $E(\varphi)$ denote the set of probability measures $\hat{\mu}$ which satisfies the condition (). The \textit{Evolutionary Entropy } $H$, associated with the potential $\varphi$ is defined by
$$
H=H_{evol}(\varphi)=\sup\{H_\mu(\tau):\mu\in E(\varphi)\}\leqno(11)
$$
The evolutionary entropy is contingent on the function $\varphi$ which describes the interaction.

\bigskip
\textbf{(3.3) Evolutionary Entropy and modes of Interaction.}
 
\smallskip
The evolutionary entropy is determined by the  potential, $\varphi$, that is the mode of interaction between the components that define the network. The nature of the interaction, local or long--range, will generate different expressions for the evolutionary entropy. We will illustrate the relation between the nature of the interaction and evolutionary entropy by considering a potential function of the form
$$
\varphi(x)=\log a_{x_0x_1}\leqno(12)
$$
Here $A=(a_{ij})$ denote the adjacency matrix of the weighted directed graph. The matrix $A=(a_{ij}) \ge 0$ is assumed to be irreducible.

\ms
The evolutionary entropy $H$  associated with the interaction  $\varphi(x)=\log a_{x_0x_1}$, Demetrius (2013), is the entropy of the Markov chain $P=(p_{ij})$, where
$$
p_{ij}=\frac{a_{ij}u_j}{\lambda u_i}\leqno(13)
$$
The element $\lambda$ is the dominant eigenvalue of $A=(a_{ij})$,  and $\bar{u}=(u_i)$ is the eigenvector associated with $\lambda$.  

\ms
Let $\pi=(\pi_i)$ denote the stationary distribution of $P$. Evolutionary entropy, $H$, is now given by
$$
H=-\sum^d_{i,j=1}\pi_{i}p_{ij}\log p_{ij}\leqno(14)
$$

The parameter $P(\varphi)$, is defined by (9), can be explicitly computed, Demetrius (2013). $P(\varphi)$ is given  by
$$
r(\varphi)=\log \lambda\leqno(15)
$$
The mean energy, $\Phi=\int \varphi d\mu$, is given by
$$
\Phi=\sum^d_{i,j}\pi_ip_{ij}\log a_{ij}\leqno(16)
$$
We have from (14), (15) and  (16), the identity
$$
\log\lambda=H+\Phi\leqno(17)
$$
The quantity $H$, given by Eq. (14) can be expressed by the relation, Demetrius and Gundlach (2014)
$$
H=\frac ST
$$
The quantity $T$ is the mean cycle time of the Markov process. The quantity $S$ , the conformational entropy, is a measure of the number and diversity of interaction cycles generated by  the directed graph. The quantities $T$ and $S$ can be  described as follows:

\ms
We write, as indicated in the previous section $X=\{(1,2,\ldots,d)\}$. This defines the nodes of the graph. We now fix an arbitrary vertex $\alpha$ in $X$, and we denote by $X^*$, the set of directed paths, which starts at $\alpha$ and ends at $\alpha$ without traversing  the node $\alpha$ in the middle. An element $\tilde{\alpha}$ is a sequence,
$$
\alpha\to \beta_1 \to \beta_2\to \cdots \to \beta_{n-1}\to \alpha
$$ 
which we denote by
$$
\tilde{\alpha}=[a \beta_1 \beta_2 \cdots  \beta_{n-1} a].
$$
Consider the function 
$$
p_{\tilde{a}}=p_{\alpha\beta_1}p_{\beta_1\beta_2}\cdots p_{\beta_{n-1}\alpha}
$$ 
The quantity $p_{\tilde{\alpha}}$ denote the probability of a randomly chosen cycle which begins at $\alpha$  and ends at $\alpha$.

\ms
The quantity $T$, the cycle time,  and the quantity $S$, the conformational entropy, are: 
$$
T=\sum_{\tilde{\alpha}\in X^*}|\tilde{\alpha}|p_{\tilde{a}}\, ; \quad  S=\sum_{\tilde{\alpha}\in X^*}p_{\tilde{a}}\log p_{\tilde{a}}\leqno(18)
$$
The application of Directionality Theory  to the study of  self--organization is based on the relation between  the evolutionary entropy,  and the following properties  which describe the behavior of the network, Demetrius (2013). 
\begin{enumerate}
	\item[(a)] \textit{Cooperativity:} The property refers to the extent to  which the components  that comprise the network engage in pairwise reciprocal interactions. Evolutionary entropy is a statistical measure of cooperativity.
\item[(b)] \textit{Stability:}   This property describes the rate at which the population returns to the steady state  after a random perturbation of the interacting   components.  Evolutionary entropy is positively correlated with  stability.
\item[(c)] \textit{Darwinian Fitness:} The term fitness, in this context describes the rate at which the population converts the external resource  endowment into metabolic energy, and biological work. Evolutionary entropy is a statistical measure of Darwinian fitness.
\end{enumerate}

\bigskip
\textbf{(3.4) Evolutionary   Entropy and Cooperativity}

\smallskip
The concept cooperativity describes the extent to which the units that comprise the network engage in acts of reciprocity in the sharing and distribution of energy. Cooperativity can be formally described by the number of dyadic interactions between the nodes of the graphs. This property is represented  by the evolutionary   Entropy.

\ms
Networks described in terms of increasing values of evolutionary   entropy $H$  are given by Fig. 5(a), (b), and (c).

\ms
The network in Fig. 5(a) assumes its minimum value, namely $S=0$.

\ms
The network in Fig. 5(c) assumes its maximum value, $S =\log 3$.

\ms
\begin{center}
	\includegraphics[width=10cm]{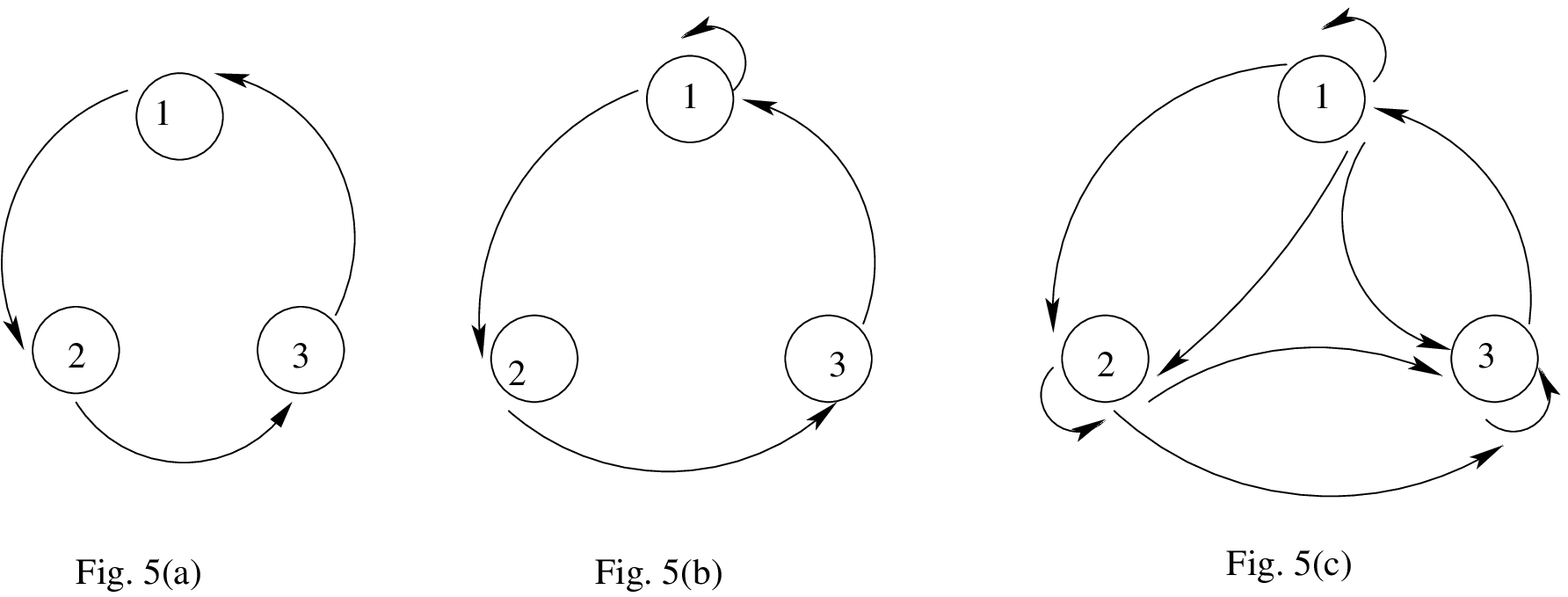}
	
\end{center}

\bigskip
\textbf{(3.5)   Evolutionary Entropy and Stability}

\smallskip
The stability of the network is defined as the rate at which macroscopic properties of the network return  to the steady state condition after a perturbation. 
The mathematical formalism of large deviation theory offers an analytic framework for representing  the stability of the network. 

\ms
The  network can be described by the mathematical object $(\Omega,\mu,\varphi)$. The function $\varphi:\Omega\to R$, the potential, describes the local interaction between the individual components. The probability measure $\mu$ describes the extent to which the energy is distributed among the components of the network. 

\ms
The mean energy is given by the function $\Phi$ defined by
$$
\Phi=\int \varphi d\mu
$$
Write
$$
S_n\varphi(x)=\sum^{n-1}_{j=0}\varphi(\tau(x))
$$
Let 
$$
P_n(\varphi) = |\frac 1n S_n(\varphi(x))-\Phi|\leqno(19)
$$
The quantity $P_n(\varphi)$ represents the deviation of the sample mean $\frac 1n S_n(\varphi(x))$ from the normalized mean value.

\ms
Let $Q_n(\ve)$ denote the probability that the sample mean $\frac 1n S_n\varphi(x)$, differs from the mean value $\Phi$,  by more than $\ve$.  The rate at which the system returns to the steady state value after a random perturbation is given by
$$
R=\lim\limits_{\ve\to 0}\lim\limits_{n\to\infty}\left[\frac 1n\log  (Q_n(\varepsilon))\right]
$$
The quantity $R$, denoted Robustness, is a measure of the relaxation time of the system after a random perturbation.

\ms
In Demetrius (2013), Demetrius and Gundlach (2015), we invoked the theory of large deviations to show that the stability parameter, $R$, and the  evolutionary entropy, $H$, are positively correlated. A formal expression of the correlation can be elaborated as follows:

\ms
We  consider a perturbation of the potential $\varphi$. We assume that the perturbation has the form 
$\varphi(\delta)$, which is given by
$$
\varphi(\delta)=\varphi+\delta\varphi
$$
Let $\mu(\delta)$ denote the equilibrium measure associated with the function $\varphi(\delta)$, and let $H(\delta)$ denote the perturbed entropy.

\ms
The changes $\Delta R$ and $\Delta H$ are defined by
$$
\Delta R = R(\delta) - R(0)
$$
$$
\Delta H = H(\delta) - R(0)
$$
The Complexity--Stability Theorem, Demetrius (2013), Demetrius and Gundlach (2015), asserts that evolutionary entropy $H$, and robustness $R$, are positively correlated, that is
$$
\Delta H.\Delta R > 0\leqno(20)
$$
for sufficiently small $\delta$.

\bigskip
\textbf{(3.6) Evolutionary Entropy and Darwinian Fitness}

\smallskip
The concept Darwinian fitness describes the rate at which the population transforms the external energy source into metabolic energy and biological work. 

\ms
The process which describes the transformation of the resource endowment into population numbers is represented by Fig. (6)

\ms
\begin{center}
	\includegraphics[width=10cm]{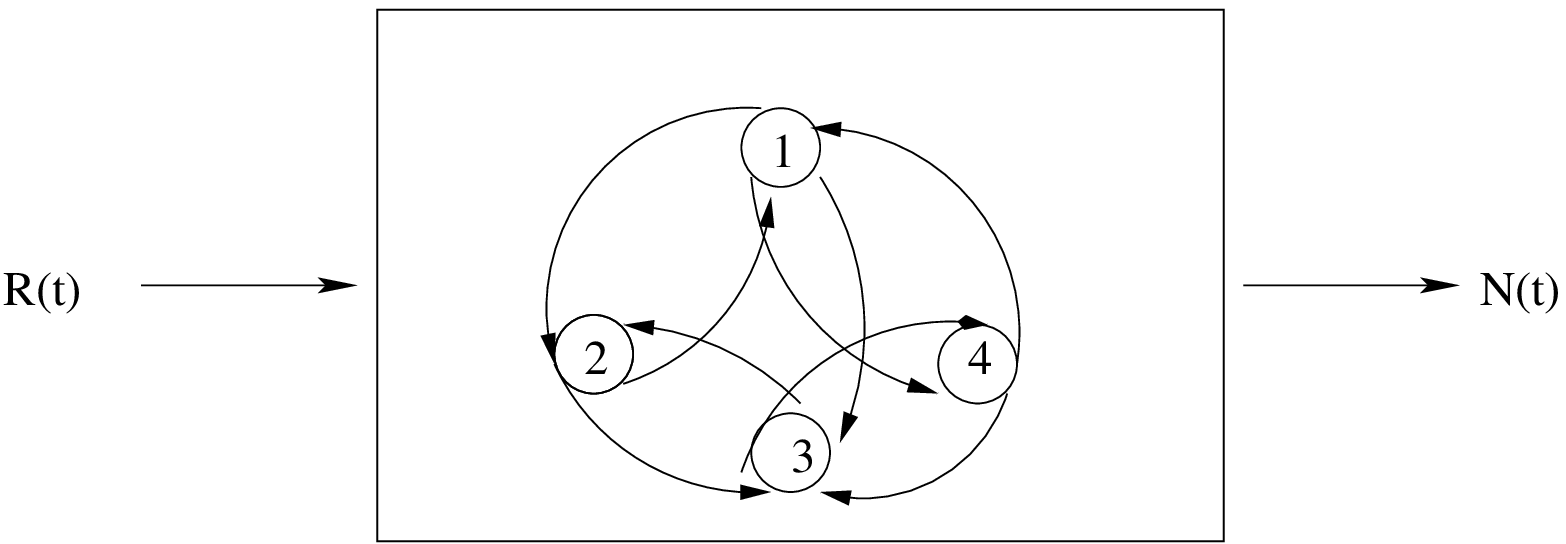}
	
	\ms
	Fig. (6): \textit{Interaction between the external resource and the population}
	
\end{center}

\ms
The quantity $R(t)$ denote the resource amplitude at time $t$, and $N(t)$ the population size.

\ms
We will show that Darwinian fitness, the  rate at which the system converts the energy generated by the resource, into metabolic energy and biological work, is analytically described by evolutionary entropy $H$.

\ms
We write
$$
H=\lim\limits_{t\to\infty} \frac 1t\log \left[\frac{N(t)}{R(t)}\right]\leqno(21)
$$

\ms
To establish (21), we assume that the Resource process and the Population process are in dynamical equilibrium.

\ms
If $\psi$ denote the resource production rate, and $r$ denote the population growth rate, then the assumption that the two processes are in dynamic equilibrium entails
$$
\psi=\frac{dr(\delta)}{d\delta}|_{\delta=0}\leqno(22)
$$
The function $r(\delta)$ is the population growth rate of the dynamical system $(\Omega,\mu(\delta),\varphi(\delta))$. The function  $\varphi(\delta)$, a perturbation of the system $(\Omega,\mu,\varphi)$, is defined by
$$
\varphi(\delta)=\varphi+\delta\varphi$$

\ms
Now, a  perturbation analysis, Demetrius (2013), shows that
$$
\frac{dr(\delta)}{d\delta}\Bigg|_{\delta=0} = \Phi
\leqno(23)$$
where $\Phi$ denote the reproductive potential of the system $(\Omega,\mu,\varphi)$.

\ms
We conclude from (22) and (23) that, when the resource process and the population process are in dynamical equilibrium, the reproductive potential  and the resource production rate coincide. We write
$\Psi=\Phi$, to express the identity.

\ms
In view of this identity, the  evolutionary entropy $H$ is now given by
$$
H=r-\Psi\leqno(24)
$$
Since 
$$r=\lim\frac 1t\log[N(t)], \, \, \mbox{\rm and}\, \, 
 \psi=\lim\frac 1t\log [R(t)],
$$
we conclude that the rate at which the population converts the resource endowment $R(t)$ into  biological work is given by evolutionary entropy.

\newpage

\centerline{\large\bf{4. Collective  Behavior: Directionality Theory and Self--Organization}}

\ms
The evolutionary dynamics of the collective behavior of macroscopic aggregates at various scales of organization --- molecular, cellular, multicellular --- can be analyzed in terms of the statistical parameter \textit{evolutionary entropy, } a measure of the cooperativity of the interacting  components.

\ms
Evolutionary entropy, a measure of the distribution of the internal energy of the system among the components,  describes the rate at which the components transform the external energy into mechanical energy and biological work.

\ms
Directionality Theory is the study of changes in evolutionary entropy  subject to the processes: Variation, succession  and natural selection. The theory of collective behavior was developed to provide an abstract mathematical model of the Darwinian theory of evolution. Variation in the Darwinian  context pertains to mutation --- random changes in the genome. Genetic mutations  induce irreversible changes in morphology, physiology or behavior. Succession refers to genetic inheritance.

\ms
The  directional changes in evolutionary entropy due to the process of variation and natural selection is qualitatively annotated in the following tenet.

\ms
\textit{The Entropic  Principle of Evolution:} The outcome of competition between an incumbent and a variant population is contingent on the amplitude and variation of the external energy source, and the population size. The changes are  characterized by extremal states of evolutionary entropy .

\ms
Analytically, the principle asserts that the steady state of the variation--selection process will be extremal states of evolutionary entropy. These states satisfy the condition
$$
(-\Phi+\gamma/M)\Delta H \ge 0\leqno(25)
$$
The parameters $\Phi$ and $\gamma$ describe the resource production rate, and its temporal correlation, respectively. The quantity $M$ denote the population size, and $\Delta H = H^* - H$, where $H$ and $H^*$ are the evolutionary entropy of the incumbent and variant population, respectively.

\ms
The theory of self--organization developed in this article is the study of changes in evolutionary entropy in macroscopic aggregates subject to the forces of variation and selection.  Variation in the framework  of self--assembly  pertains to fluctuations in the interaction of the components. These fluctuations are engendered by  the lability of the interactions --- a condition which induces reversible changes in the physical, chemical or behavioral properties of the components.  The main proposition  of the theory of Self--Organization is  the following rule:

\ms
\textit{The Entropic Principle of Self--Organization:} The equilibrium states of self--organizing processes are  configurations which maximize evolutionary entropy, contingent on the external energy source.

\ms
Analytically, the Principle asserts that the steady state of the process of fluctuation and selection will be given by the relation
$$
-\Phi\Delta H \ge 0\leqno(26)
$$
The function $\Phi$ refers to the external energy source, whereas $\Delta H$ is given by $\Delta H = H^*-H$.

\ms
The mathematical basis for the Entropic Principle of Evolution, as formalized by (25),  is described in Demetrius (1983), Demetrius and Gundlach (2015). We will review the main ideas underlying the derivation of (25) in Section (4.1).

\ms
The Entropic Principle of Self--Organization can be derived from (25) by imposing the condition $\gamma=0$. In Section (4.3), we will provide the  rationale for the condition $\gamma=0$.

\bigskip
\textbf{(4.1) Directionality  Theory and Collective Behavior}

\smallskip
The process generated by the directed graph, Fig. (1),  can be represented by the mathematical object $(\Omega,\mu,\varphi)$, where:
\begin{enumerate}
	\item[(i)] $\Omega$: The set of genealogies, that is the set of paths generated by the graph
	\item[(ii)]	$\varphi$:  A potential function on $\Omega$ which describes the interaction between the individual components.
\item[(iii)]	$\mu$: A probability measure on $\Omega$, which describes the distribution of energy between the components of the network.	
	\end{enumerate}
 The macroscopic parameters that describe the population dynamics of the network are \begin{enumerate}
	\item The population growth rate, $r$, where $r=P(\varphi)$, as defined by (9)
	\item The demographic variance $\sigma^2$, which is given by
$$
\sigma^2=\int(\varphi-\int\varphi d\mu)^2d\mu
$$
\end{enumerate}
We assume that the  behavior of the incumbent  with population size $N(t)$  is described by the dynamical system $(\Omega,\mu,\varphi)$. 

\ms
The variant with population size $N^*(t)$ is defined by the dynamical system $(\Omega, \mu(\delta), \varphi(\delta))$,  where the perturbation $\varphi(\delta)$ is given by
$$
\varphi(\delta)=\varphi + \delta\varphi
$$
Let $f(N(t)$ and $f^*(N^*(t))$ denote the density of the processes $N(t)$ and $N^*(t)$. Let $(r,\sigma^2)$ and 
$(r^* ,\sigma^{*2})$ denote the growth rate and demographic variance of the incumbent population, $(\Omega,\mu,\varphi)$, and the variant population $(\Omega,\mu(\delta),\varphi(\delta))$, respectively.

\ms
The evolution of the densities $f(N,t)$ and $f(N^*,t^*)$  is given by the solution of the Fokker--Planck equation
$$
\frac{\partial f}{\partial t}=- r\frac{\partial(fN)}{\partial N}+\sigma^2\frac{\partial^2(fN)}{\partial N^2}
$$
and
$$
\frac{\partial f^*}{\partial t} = - r^*\frac{\partial(f^*N^*)}{\partial N^*} + \sigma^{*2}\frac{\partial^2(f^*N^*)}{\partial N^{*2}}
$$
Let $y$ denote the initial frequency of the variant population, and $P(y)$ the probability that the diffusion process leads to an absorption in state 1 --- which corresponds to extinction of the population. We obtain, see Demetrius and Gundlach (2000), 
$$
P(y)=\frac{\ds 1-\left(1-\frac{\Delta\sigma^2y}{\sigma^{*2}}\right)^{\frac{2M_s}{\Delta\sigma^2}+1}}{\ds 1-\left(1-\frac{\Delta\sigma^2}{\sigma^{*2}}\right)^{\frac{2M_s}{\Delta\sigma^2}+1}}\leqno(27)
$$ 
where
$$
s=\Delta r - \frac 1M\Delta \sigma^2\leqno(28)
$$	
The probability $P(y)$ is determined by its convexity, and the size of $M$. The convexity of $P$ can be expressed in terms of $s$ only.

\ms
The analysis of the first and second derivatives of the function $P(y)$ enables the characterization of the convexity of $P(y)$ in terms of the sign of $s$. We obtain, see Demetrius (2015)

\ms
$$s > 0 \Longrightarrow P(y)\, \, \mbox{\rm convex}
$$
$$s < 0 \Longrightarrow P(y)\, \, \mbox{\rm concave} 
$$
The relation between $s$ and the geometry of $P(y)$ indicates that the outcome of competition between the incumbent and the variant will be determined by $\Delta r$ and $\Delta \sigma^2$. We can infer that the selective advantage, the condition which specifies the invasion or extinction of a variant is given by
$$
s = \Delta r - \frac 1M\Delta\sigma^2
$$
We can express the outcome of competition between the incumbent and variant  population uniquely in terms of the change in entropy $\Delta H$.

\ms
Now, a perturbation analysis of the evolutionary entropy $H$ shows that
$$
\frac{dH(\delta)}{d\delta}\bigg|_{\delta=0} =- \sigma^2
$$
Hence $\Delta H \approx -\sigma^2\delta$. Write
$$
 \Phi=\frac{dr(\delta)}{d\delta}\bigg|_{\delta=0}\, \, ; \, \, \gamma =  \frac{d\sigma^2(\delta)}{d\delta}\bigg|_{\delta=0}
$$
This yields
$$
\Delta r \approx \Phi\delta\,, \,  \Delta \sigma^2\approx\gamma\delta
$$
Write
$$\tilde{s}=(-\Phi+\gamma/M)\Delta H\leqno(29)$$
We obtain, using the expressions for $\Delta r$, $\Delta\sigma^2$ and $\Delta H$, that $s\tilde{s} > 0$.

\ms
We conclude that $\tilde{s}$ is also a measure of selective advantage. Hence the change in evolutionary entropy  due to natural selection is: 
$$
(-\Phi+\gamma/M)\Delta H \ge 0\leqno(30)
$$
The relation (30), qualitatively, asserts that the outcome of competition between an incumbent and a variant population is contingent on the statistical parameters $\Phi$, $\gamma$, and the population size $M$, and characterized by extremal states of  evolutionary entropy. 

\ms
The parameters $\Phi$ and $\gamma$ are related to the forces which provide the energy,  internal and external, that regulate  collective behavior.

\ms
\textbf{(4.2)  Self--Organization and Collective Behavior}

\smallskip
Self--organization is the study of changes in the evolutionary entropy of a macroscopiuc aggregate subject to the forces of fluctuation and selection.

\ms
The parameters that define the external energy resource are the functions $\Phi$ and $\gamma$ defined by
$$
\Phi=\frac{dr(\delta)}{d\delta}\Bigg|_{\delta=0}\, \, ;\, \, \gamma=\frac{d\sigma^2(\delta)}{d\delta}\Bigg|_{\delta=0}
$$
The variation  induced in processes  of self--assembly derive from the lability of the interactions between the components. these changes which we define as fluctuations are reversible. Hence  $\gamma=0$.

\ms
The steady state of the process of fluctuation and selection will now be given by the relation
$$
-\Phi\Delta H \ge 0\leqno(31)
$$
A qualitative expression of the relation (31) now follows:

\ms
\textit{The Entropic Principle of Self--Organization:}  The equilibrium states of self--organizing processes are the configurations which maximize evolutionary entropy, contingent on the external energy source.

\ms
We will apply the relation  (31) to derive general rules which  encode the emergence of order in both static self--assembly, and dynamic self--assembly.

\ms
The identity
$$
r=H+\Phi\leqno(32)
$$
 enables a formal distinction between the two modes of self--assembly, static and dynamic.

\ms
\textbf{(I)} \textit{Static Self Assembly}

\smallskip
Static self assembly organizes information encrypted in the individual components into patterns whose stability to maintained uniquely by the action of internal constraints

\ms
Examples of this mode of self--organization are:
\begin{enumerate}
	\item[(i)] The folding of simple proteins with two--state behavior
	\item[(ii)] The assembly of molecular crystals
	\item[(iii)]The formation of liquid crystals.
\end{enumerate}

The local equilibrium which defines this process of self--assembly entails that the growth rate, $r$,  satisfies the condition $r=0$.

\ms
We obtain from (32) the relation
$$
H+\Phi=0\leqno(33)
$$
In view of (33), the condition (31) reduces to the relation 
$$
S.\Delta H \ge 0
$$
Since $S > 0$, we conclude that the process of static self--assembly is  determined by the condition
$$
\Delta H > 0\leqno(34)
$$
The condition asserts that the process of  static self assembly manifests configurations which maximize evolutionary entropy $S$.

\ms
\textbf{(II)} \textit{Dynamic Self Assembly}

\smallskip
In dynamic self assembly, the interactions which induce  the  the stability  and organization of patterns  are determined  by  a dissipative energy source.

\ms
Dynamic self--assembly is well documented  in microtubular morphogenesis . Tubular and actin polymers determine cell shape and polarity. These polymers provide an internal structural framework for processes such as cell division.

\ms
This mode of self--organization involves a dissipative    energy source. Tubulin is an enzyme that binds and hydrolyzes GTP to GDP during assembly. The existence of a constant energy source entails that self--assembly in these system is constrained by the condition $r\neq 0$.

\ms
The relation (32) now implies that
$$
\Phi = r-H
$$
The change $\Delta H$ in evolutionary entropy, due to interaction between the individual components, implies the following relation
$$
(H-r)\Delta H > 0\leqno(35)
$$
This relation asserts that the process of dynamic self--assembly manifests configuration which maximize evolutionary entropy, contingent on the relation $H > r$, the evolutionary entropy exceeding the energy production rate.

\newpage
\centerline{\large\bf{5. Self--organization: Static and Dynamic Self--Assembly}}

\smallskip
The model of self--organization proposed in this article is based on the notion that the emergence of organized structures is the outcome of a hierarchical process, which begins with structures which are local in configuration,  and marginal in stability.  The increase in complexity is driven by a 	variation--selection process, which  can be analyzed in terms of the statistical parameter evolutionary entropy.

\ms
We will illustrate the general principles which regulate the phenomenon of self--organization by the analysis of  the two canonical modes of self--assembly, static and dynamic.

\ms
Static self--assembly will be illustrated by the folding mechanism of small proteins with two--state folding behavior. Dynamic self--assembly will be described by the emergence of Benard Convection cells.

\ms
\textbf{(5.1) Static Self--Assembly --- The Protein Folding Problem:} 

\smallskip
The protein folding problem is now understood in terms of three related issues, Dill et al. (2008)
\begin{enumerate}
	\item[(i)]  \textit{The folding code:} The encoding of the native structure of a protein in terms of the physico--chemical properties of the polypeptide chain.
	\item[(ii)]  \textit{The folding mechanism:} The dynamical system which specifies the transition from a polypeptide chain to a unique stable conformation.
	\item[(iii)] \textit{The folding principle:	} The physico--chemical principle which encodes  the transition from the amino--acid sequence to the native structure.
\end{enumerate}
A major contribution to the resolution of the protein folding problem was made by Anfinsen (1973). Experimental studies of the renaturation of ribonuclease led to the notion that the three--dimensional structure of a protein is specified completely by the amino--acid sequence, and that the native structure correspond to global free energy minima.

\ms
This empirical observation  was formalized in terms of what is now called:

\ms
\textbf{The Thermodynamic Hypothesis:} The native structure of a protein at the proper environmental conditions, temperature, solvent concentration and composition, is a unique stable and kinetically accessible minimum of the free energy.

\ms
The Thermodynamic hypothesis  is a  derivative of  the Statistical Thermodynamics  of collective behavior of amino acids. The measure of the cooperativity invoked in this model is the statistical measure, thermodynamic entropy. 
This  measure of molecular organization ignores the way in which the instantaneous configuration of the amino acids influences the behavior of the other molecules in the protein. This influence is negligible in systems described by a huge number of interacting molecules. However, the influcence becomes significant in systems where $N$, the number of degrees of freedom is large.
\ms
The relatively small size of proteins, and the long--range nature of the interatomic forces between the molecules in the polypeptide chain, entail that the collective behavior of these systems cannot be effectively described in terms of the thermodynamic formalism of Gibbs and Boltzmann. 
Empirical and computational studies of protein folding do not provide support for the  \textit{Thermodynamic hypothesis,} Sorokina and Mushgian (2018).

\ms
The failure of the \textit{Thermodynamic Hypothesis  }  to provide a  valid rationale for the  process which drives a polypeptide chain to a unique stable configuration, derives from the  ineffectiveness of thermodynamic entropy as a  measure of the cooperativity of interacting amino acids in a polypeptide chain.

\ms
We will appeal to the statistical measure, evolutionary  entropy, a generalization of thermodynamic entropy,   to propose a new principle --- the  \textit{Evolutionary  Entropic Folding Principle} --- to describe the transformation of a polypeptide chain to a 3--dimensional stable structure.

\ms
\textbf{The  evolutionary Entropic Folding Principle:} 
The native structure of a protein at the proper environmental conditions, temperature, solvent concentration and composition, is the unique stable, kinetically accessible state which maximizes evolutionary entropy, contingent on the external energy constraints.

\ms
The evolutionary Entropic Folding Principle asserts that the  emergence of spatio--temporal order in a polypeptide chain is  contingent on  evolutionary entropy. The unique stable state is determined by the condition 
$$
\Delta H \ge 0\leqno(36)
$$
The static self--assembly process which the expression (36) describes is a hierarchical process which we delineate as follows:

\ms
The disordered state of a protein consists of individual segments of the polypeptide chain that move relative to one another; and groups that rotate about a single bond. This state has many of its hydrophobic side chains exposed to solvents.  The disordered state is defined by an inherently small value for the evolutionary entropy.

\ms
The native state of the protein has many of its hydrophobic side chains shielded from water because they are packed in hydrophobic cores. This packing entails an enhanced pairwise contact between interacting amino--acid residues. This enhancement in contact entails a large evolutionary entropy.  

\ms
The transition from the unfolded state to the three dimensional structure proceeds by the formation of local structures of marginal stability. These local structures interact to form intermediates with higher contact order, and increased stability. The equilibrium states are the configurations that maximize the  number of pairwise contacts. These states are characterized by maximal evolutionary  entropy.

\ms
The folding principle  (7)  refers to the two state  foldings of small proteins. Folding in these proteins occur without dissipating energy, and  can be  characterized as processes of  static self--assembly.

\ms
\textbf{(5.2) Dynamic Self--Assembly --- Benard Convection:}  

\smallskip
The emergence of spatio--temporal order  requires a continuous input of energy. The emergence of a stable macroscopic pattern is now contingent on a relation between the evolutionary  entropy  and the growth rate.   The transition from a disordered system to an ordered macroscopic structure is now given  in terms of the following condition,
$$
(H-r)\Delta H \ge  0\leqno(37)
$$

\ms
Benard Convection is one of the classical examples of dynamic self--assembly. The phenomenon is represented by the  emergence of macroscopic patterns in a horizontal fluid layer heated from below.

\ms
The statistical parameter, evolutionary entropy, is a measure of the geometry of convection patterns. The  disordered state, the initial condition, is described by a small value for evolutionary entropy.   
The changes in  evolutionary  entropy are due to the temperature gradient, induced by the external source of energy, and the gravitational force, an internal constraint. For small values of this gradient, the outcome is a transfer of heat by conduction in a fluid at rest. As the gradient increases, the local structures interact to generate intermediates with increased evolutionary  entropy,  and  enhanced stability.

\ms
At a critical value of the temperature gradient, the  evolutionary entropy exceeds the growth rate, a measure of the opposing gravitational force. 

\ms
Below instability, $H < r$,  the energy of the system is distributed in the random thermal motion of the molecules. Beyond instability $H > r$, the energy of the system is expressed as an organized macroscopic pattern. This stable arrangement of the cells correspond to the state which maximizes evolutionary entropy. 

\newpage 

\centerline{\large\bf 6. Conclusion}

\ms
Directionality Theory, is the study of dynamical changes in the collective behavior of populations of interacting components subject to the processes of variation and   natural selection. 

\ms
The theory is structured in terms of the statistical parameter evolutionary entropy, a measure of   cooperativity --- the extent to which the internal energy of the system is shared and distributed between the interacting components. Evolutionary entropy also describes the rate at which the components convert the energy of the external resources into metabolic energy and work. 

\ms
The dynamical changes in collective behavior is contingent on the nature of the mechanism which drives the process of variation.  We will distinguish between  two mechanisms of variation, namely,  genetic mutation, a process which  defines Darwinian evolution;  and   fluctuation, a phenomenon which characterizes self--assembly.

\bigskip
\textbf{6.1 Variation by Mutation: The Evolutionary Process}

\smallskip
Mutations generate irreversible changes in the cells or the organisms that comprise the population. The irreversibility of the variation process entails that statistical parameters $\Phi$ and $\gamma$ which characterize the energy source, satisfy the condition $\Phi\neq 0$, $\gamma\neq 0$.

\ms
Hence the change $\Delta H$ in evolutionary entropy under the mutation--selection process will be described by the relation
$$
(-\Phi+\gamma/M)\Delta H \ge 0
$$
This condition implies that evolution will result in \textit{extremal} states of entropy entropy, contingent on the constraints on the energy source. Since evolutionary entropy is positively correlated with structural stability of the system, we can infer that the process may result in an increase or decrease in stability.

\bigskip
\textbf{6.2 Variation by Fluctuation: The Self--Organization Process}

\smallskip
Fluctuations generate reversible changes in the components that comprise the population. The reversibility of the variation process implies that the statistical parameter  $\gamma$, which descrivbes the correlation in the resource process satisfies the condition $\gamma = 0$.

\ms
The constraint on $\gamma$ entails that the change in evolutionary entropy will be described by the relation
$$
-\Phi\Delta H \ge 0$$
Since the energy source in Self--Organizing processes, satisfy the condition $\Phi \le 0$, we can infer, that the self--organizing process will always result in an increase in evolutionary entropy, and concomitantly an increase in stability.

\ms
\textbf{6.3. The Steady State: Self--Organizing and Evolutionary Processes.}

\smallskip
Evolutionary entropy, a statistical measure of cooperativity, and stability, the  rate at which the population returns to its steady state condition after a perturbation, are positively correlated. This relation is pertinent in distinguishing  between the equilibrium states of Self--organizing and evolutionary processes.

\ms
The equilibrium states of Self--Organizing processes are states which maximize evolutionary entropy. Accordingly, the equilibrium states of these processes will be stable.

\ms
The equilibrium states of evolutionary processes are extremal states of evolutionary entropy. The nature of these equilibria, maxima or minima of evolutionary entropy, will be constrained by the energy source that drives the process.  Consequently,  these states will be stable or unstable, contingent on the resource constraints .

\ms
Accordingly,  the steady states of Self--organizing processes are characterized by their stability, that is, their resilience in the face of perturbation. However, the steady states of evolutionary processes are characterized by their adaptation to the external environment.

\newpage
\centerline{\large\bf  References}
\begin{enumerate}
	\item Anfinsen, C.B. (1973): Principles that govern the folding of protein chains. \textit{Science} 181, 223--230.
	\item Arnold, L., Demetrius, L. and M. Gundlach (1994): Evolutionary Formalism for products of positive random matrices. \textit{Ann. Appl. Probab.} 4, 859--901.
	\item Baker, D. (2000): A surprising simplicity to protein folding. \textit{Nature}, 405, 39--42.
		\item Bowen, R. (1975): Equilibrium States and the Ergoidic Theroy of Anosov diffeomorphism. Springer Verlag.
\item Camazine, S. (2003): Self--Organization in Biological Systems. Princeton Studies in complexity. Princeton University Press.
\item Crooks, GE. (1999): Entropy Production Fluctuation. Theorem and the Nonequilibrium Work Relation for Free Energy Differences: \textit{Physical Review} E-60 No 2721.
	\item Davies, P., Rieper, E. and J. Tuszynski (2012): Self--Organization and Entropy reduction in a living cell. \textit{Biosystems }Vol. 111, 1--10.
	\item Demetrius, L. (1983): Statistical mechanics and population Biology. \textit{Jour. Stat. Phys.} Vol. 30, 709--753. 
	\item Demetrius, L. (1974):	Demographic parameters and natural selection. \textit{Proc. Natl. Acad. Sci.} 4645--4649.
	\item Demetrius, L. (1975): Natural selection and age--structured population.\textit{ Genetics} 79, 533--544.
	\item Demetrius, L. (1997):	Directionality Principle in Thermo--dynamics and Evolution. \textit{Proc. Natl. Acad. Sci.} Vol. 94, 3491--3498.
	\item Demetrius, L.; Gundlach, V.M. and G. Ochs (2009): Invasion exponents in biological networks.\textit{ Physica A.} 388--572.
	\item Demetrius, L. (2013): Boltzmann, Darwin and Directionality Theory. \textit{Physics Reports}, Vol. 530, 1--85.
	\item Demetrius, L. and Matthias Gundlach (2014): Directionality Theory and the Entropic Selection Principle.\textit{ Entropy} 15, 5428--5622.
	\item Demetrius, L.	 and C. Wolf (2022): Directionality Theory and the Second Law of Thermodynamics, \textit{Physica A., Statistical Mechanics and its Applications. }Vol 598, 1273e5
	\item Dill, K.A., Ozkan, S.B., Shell, M.S. and T.R. Weikl (2008): The Protein Folding Problem. \textit{Annual Review of Biophysics} 37, 289--336.
	\item Germano, F. (2022): Enropy directionality Theory and the evolution of Income inequality. \textit{Jour. of Economic Behavior and Organization.} Vol. 198, 15--43.
\item Haken, H. (1977): Non--equilibrium phase transitions and self--organization in physics, chemistry, and biology in Synergetics: An Introduction. Springer, Berlin.
\item Jarzynski, C. (2011): Equalities and Inequalities: Irreversibility and the Second Law of Thermodynamics. \textit{Ann. Rev. Condens. Matter Phys.} 2: 329--351.
\item Karsenti, E. (2008): Self--organization in cell Biology: a brief history. \textit{Nat. Rev. Mol. Cell. Biol.} 9, 255--262.
\item	Kirschner, M.  and T. Mitchison (1986): Beyond Self Assembly: From Microtubules to Morphogenesis. \textit{Cell}, Vol 45, 329-342.
\item Krugman, PR. (1996):L The Self--Organizing Economy. Cambridge, Mass. 
\item Lehn, J.M. (2012): Constitutional Dynamic Chemistry: Bridge from Supramolecular Chemistry to Adaptive Chemistry. \textit{Top. Curr. Chem.} 322, 1--32. 
\item Li, Y, and M.E. Cates (2020): Steady state entropy production rate for scalar Langevin field theories. \textit{Journal  Statistical Mechanics. }
\item Morowitz, H. and E. Smith (2007): Energy flow and the Organization of Life.\textit{ Complexity }13, 51--59.
\item Nicolis, G. and I. Prigogine (1977): Self--organization in Non--equilibrium systems: From Dissipative Structures to Order through fluctuation. Wiley.
\item Prigogine, I (1969): Structure, dissipation and Life. In: Theoretical Physics and Biology ed. M. Marois, Amsterdam.	
\item Ruelle, D. (1978): Thermodynamic Formalism, Addison--Wesley Reading MA.
\item Saha, T. and M. Galic (2018): Self--Organization across scales: from molecules to organisms.\textit{ Phil. Trans. R. Soc. B.} Vol. 373: 20120118
\item Shakhnovich, E. (1994): Theoretical studies of protein folding thermodynamics and kinetics. \textit{Current Opinion in Natural biology}, Vol. 7, 27--40. 
\item Sorokina I.l and A. Mushagian (2018): Modeling protein folding in vivo  \it{Biology Direct.} Vol. 13
\item Sorokina I.l and A. Mushagian (2022): Is Protein folding a thermodynamically unfavorable, active, energy--dependent process? \textit{International Jour. Molecular Scienes,} Vol. 23, 521
	
\item Tabony, J., Glade, N. and J. Demongeot (2022): Microtubule self--organization: a biological example of emergent phenomena in a complex system.\textit{ Recent. Res. Devel. Biosphys. Chem.} 3, 11--53.
\end{enumerate}
\end{document}